\documentclass{raa}            

\usepackage{graphicx,times}             
\usepackage{natbib}
\usepackage{amssymb,amsmath}
\bibpunct{(}{)}{;}{a}{}{,}

\usepackage[a4paper=true,pagebackref=true]{hyperref}
\hypersetup{colorlinks = true, linkcolor = green, anchorcolor = red, citecolor = blue,
filecolor = red, pagecolor = red, urlcolor = red}

\usepackage{xspace}
\newcommand{\fermi}{$Fermi$-LAT\xspace}
\newcommand{\ergcms}{\,$\mathrm{erg~cm}^{-2}~\mathrm{s}^{-1}$}

\begin{document}

\title{A diagnostic of the orbital spectrum of LS 5039 with \fermi}

\volnopage{Vol.0 (20xx) No.0, 000--000}      
\setcounter{page}{1}          

\author{ Zhi Chang \inst{1}
\and Shu Zhang \inst{1,2}$^\star$
\and Yu-Peng Chen \inst{1}
\and Long Ji \inst{3}
\and Ling-Da Kong \inst{1,2}
}

\institute{
Key Laboratory of Particle Astrophysics,
Institute of High Energy Physics, Chinese Academy of Sciences,
Beijing 100049, China; $^\star${\it szhang@ihep.ac.cn}\\
\and
University of Chinese Academy of Sciences, Beijing 100049, China \\
\and 
Institut f\"ur Astronomie und Astrophysik,
Kepler Center for Astro and Particle Physics,
Eberhard Karls Universit\"at, Sand 1, 72076 T\"ubingen, Germany \\
\vs\no
{\small Received~~20xx month day; accepted~~20xx~~month day}}

\abstract{
LS 5039 is a well-known $\gamma$-ray binary system 
which consists of an unknown compact object and a massive companion O star.  
It shows rather stable emissions at high energies over years 
and hence serves as an ideal laboratory to investigate the emission mechanism 
for such peculiar systems which emit prominent $\gamma$-rays.  
To this end, we take the orbital phase resolved energy spectrum 
as observed by \fermi over 10 years. 
We divide the orbit into four orbital phases, 
each with an orbital phase range of 0.25,
centered at 0.00, 0.25, 0.50 and 0.75 respectively,
where the phase 0.0 is the periastron and phase 0.5 is the apastron.
The phases around 0.25 and 0.75 are symmetric 
and hence are supposed to have identical local acceleration environment. 
The spectral analysis shows that, 
the \fermi spectra are largely different from these two symmetric orbital phases:
the emission from orbital phase 0.25 turns out to be significantly stronger than that from 0.75. 
This result does not fit a scenario 
that $\gamma$-rays are Doppler boosted emission from bow shock tails
if LS 5039 has a shock configuration  similar to PSR B1259-63,
and indicates that the inverse Compton scatterings 
between the shock accelerated plasma and the stellar particle environment is the underline procedure.  
We also find that the previous report for a disappearance of the orbital modulation at 3--20 GeV 
is due to the similar spectral turn-over energies  of the different orbital phases.
The spectral properties of periastron and apastron regions are addressed in the context 
of the measurements in phase regions around 0.25 and 0.75. 
}

\authorrunning{Zhi Chang et al.}            
\titlerunning{SED of LS 5039 by \fermi}  

\maketitle

%
%
\section{Introduction}           
\label{sect:intro}

$\gamma$-ray binary system which consists of a compact object (neutron star or black hole)
and a high-mass OB star, is a special class of X-ray binaries (XRBs), emitting radiation in  broad  wavelengths,
from radio to high energy $\gamma$-ray band (TeV and/or GeV).
The compact stars of all $\gamma$-ray binary systems are unknown except for PSR B1259-63,
which is a non-recycled, spin-down powered radio pulsar \citep{johnston1992}.
Please see \cite{dubus2013} and \citet{dubus2015} for reviews on these $\gamma$-ray binary systems.

\begin{figure}
\centering
\includegraphics[width=0.5\textwidth]{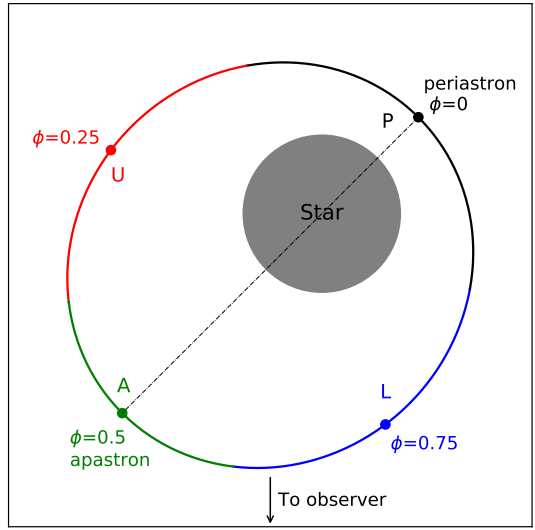}
\caption{The LS 5039 orbit as defined in \citet{casares2005}.
Relevant orbital points, such as the periastron and apastron are indicated. 
The dashed line connects the periastron and apastron.
The different colors of the orbit (P, U, A, L) show how we divided the orbit in our following analysis.}
\label{fig:orbit}
\end{figure}

LS\,5039 is known as a relatively compact binary system
for which the separation between the two objects is $\sim$0.1--0.2 AU.
The compact star is moving around an O6.5V main-sequence star with a period $P_{\rm orb} \approx 3.9$ days
and a moderate eccentricity of $0.35\pm0.04$ \citep{casares2005}. 
A schematic view of the binary system is shown in Figure \ref{fig:orbit}.

The companion of LS  5039 is an O star, which has rather stable stellar wind.
The entire system shows with emissions at multi-wavelength covering from radio to TeV band. 
The broad band emissions are very stable, as shown in \citet{takahashi2009},
and the orbital light curves at soft X-rays keep the same profile from orbit to orbit over years.
Hence this system serves as an ideal laboratory for studying the emission mechanism at work 
for $\gamma$-ray binary system, 
especially for those with compact object supposed as neutron star and hence 
the particles are accelerated via shock between stellar wind and the pulsar wind. 

The detailed multi-wavelength spectral energy distributions (SED) and orbital light curves analysis of LS 5039
have been reported in \citet{chang2016}.
At X-ray and soft $\gamma$-ray (1\,keV--30\,MeV) energies
the flux peak is located around the region of inferior conjunction (INFC),
when the compact object is in front of the massive star
(corresponding to the region of superior conjunction (SUPC),
when the compact object is behind the massive star) \citep{takahashi2009, collmar2014}.
At higher energy $\gamma$-rays, from a few hundreds of MeV to about GeV, 
the flux peak in orbital light curves moves from phase 0.6 toward 1.0 
(from around apastron to periastron) \citep{abdo2009, hadasch2012}.
Generally, the orbital light curves show a single-peaked profile with a broad peak.
Interestingly, the modulation disappears at energies of 3--20\,GeV \citep{chang2016}, 
but shows up again in the TeV band  with a flux peak located at around apastron to INFC,
quite similar to the lowest energies \citep{aharonian2006}.
However, more joint diagnostic of the source by combining the spectrum and 
the orbital phase are essential for providing more clues to our understanding of the system.

In this paper, 
we investigate in details the orbit-phase resolved energy spectrum  as observed by \fermi over 10 years. 
This paper is organized with the observations and data analysis in Section 2, 
the results in Section 3, and finally the discussion in Section 4.

\section{Observations and data analysis}
\label{sect:data}

The LAT (the Large Area Telescope) on-board $Fermi$ satellite is
an electron-positron pair production telescope operating at energies 
from $\sim$100\,MeV to greater than 300\,GeV \citep{atwood2009}.

The \fermi data included in this paper cover the period from 2008 August 4 to 2019 April 18.
The analysis of the \fermi data was performed using the  standard Science Tools
({\tt Fermitools}\footnote{\url{https://fermi.gsfc.nasa.gov/ssc/data/analysis/software/}}, version 1.0.1). 
Photons from the “P8 Source” event class (evclass=128) and “FRONT+BACK” event type (evtype=3) 
were selected\footnote{\url{https://fermi.gsfc.nasa.gov/ssc/data/analysis/documentation/Pass8\_usage.html}}.
The “P8R3\_SOURCE\_V2” instrument response functions (IRFs) were used in the analysis. 
All photons in the energy range of 0.1--300GeV and 
within a circular region of interest (ROI) of 10$^\circ$ radius centered on LS 5039 were considered.  
To reject contaminating $\gamma$-rays from the Earth’s limb, 
only events with zenith angle $\leq$90$^\circ$ were selected.

The $\gamma$-ray flux and spectral results presented in this work were calculated by 
performing a binned maximum likelihood \citep{mattox1996} fit using the tool 
{\tt gtlike}\footnote{\url{https://fermi.gsfc.nasa.gov/ssc/data/analysis/scitools/help/gtlike.txt}}. 
The known $\gamma$-ray sources within 15$^\circ$ of LS 5039, based on a LAT 8-year Point Source Catalog
(4FGL)\footnote{\url{https://fermi.gsfc.nasa.gov/ssc/data/access/lat/8yr\_catalog/}},
as well as Galactic and isotropic diffuse emission components 
(“gll\_iem\_v07.fits” and “iso\_P8R3\_SOURCE\_V2\_v1.txt”,
respectively\footnote{\url{https://fermi.gsfc.nasa.gov/ssc/data/access/lat/BackgroundModels.html}} ) 
are constructed to perform the likelihood analysis.
The spectral parameters of these sources were fixed at the values listed in the catalog, 
except for those within 3$^\circ$ of LS 5039, for which all the spectral parameters were left free.
The significance of the $\gamma$-ray fluxes coming from LS 5039 is characterized by Test Statistic (TS),
which is defined as
\begin{equation}
{\rm TS} =2(\log \mathcal{L}-\log \mathcal{L}_{0})
\end{equation}
where $\log \mathcal{L}$ and $\log \mathcal{L}_{0}$ are the logarithms of the maximum likelihood 
of the complete source model and of the null hypothesis model 
(i.e., the source model without LS 5039 included), respectively. 
The larger the value of TS, the less likely the null hypothesis is correct 
(i.e., a significant $\gamma$-ray excess lies  at the tested position) and 
the square root of the TS is approximately equal to the detection significance of a given source.

In our results, an exponentially cut-off power-law model was applied  to LS 5039 during the likelihood analysis,
which is described by
\begin{equation}
\frac{{\rm d}N}{{\rm d}E} = N_0 E^{\gamma}{\rm exp}\left(-\frac{E}{E_{\rm cut}}\right)
\end{equation}

\section{Results}
\label{sect:results}

\begin{figure}
\centering
\includegraphics[width=0.98\textwidth]{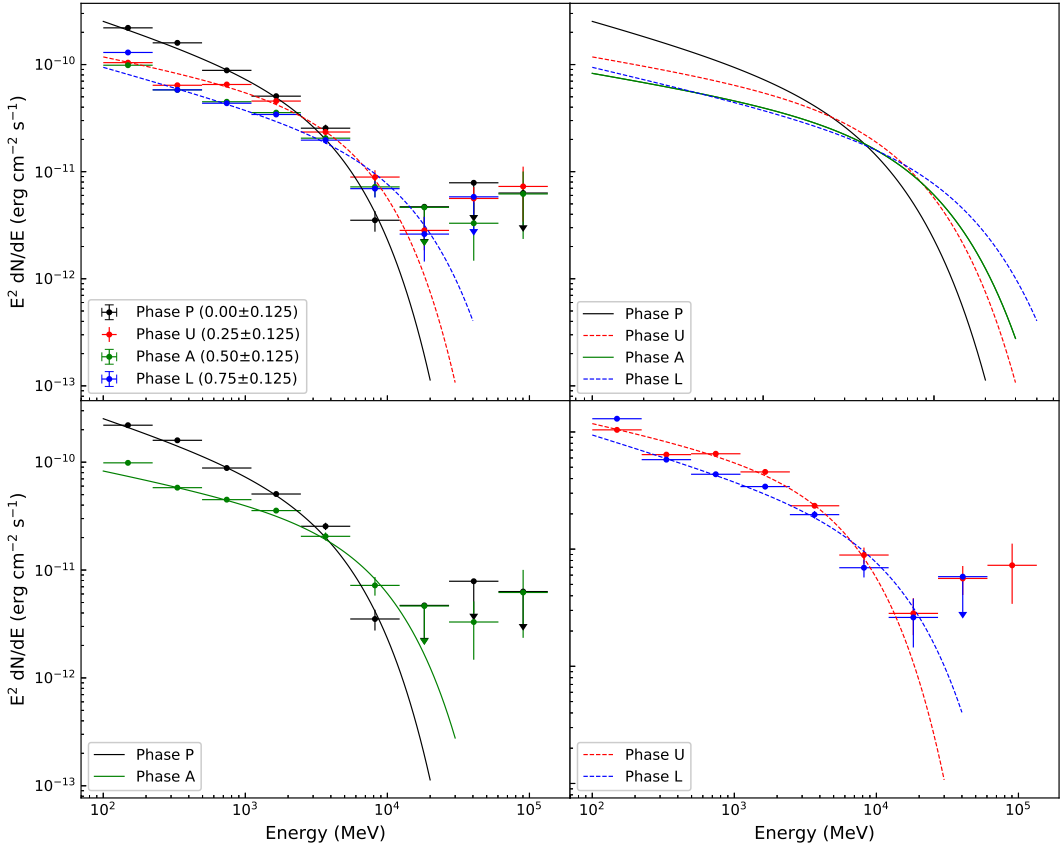}
\caption{SED derived in the likelihood analysis. 
The orbit is divided into four parts uniformly, which is described in detail in text. 
The $top$ two panels show the overall comparison of these four SEDs.
The $bottom left$ and $bottom right$  panel clearly show the comparison of the averaged spectrum 
between phase P and phase A, phase U and phase L, respectively.}
\label{fig:sed_4bin}
\end{figure}

To investigate the orbital dependence of the spectra in more detail,
we divided the orbit into four parts uniformly\footnote{
The phase in our analysis is determined by the time range and divided uniformly into four parts,
since the geometric phase is not uniform due to the change of the orbital velocity.}, 
with an orbital phase range of 0.25, centered in orbital phase 0.00, 0.25, 0.50 and 0.75 respectively
(phase P, U, A, L), as shown in different colors in Figure \ref{fig:orbit}.
The previous orbital phase selections are usually the inferior conjunction (INFC) from 0.45 to 0.9
and superior conjunction (SUPC) from 0.9 to 0.45, 
as is shown for example in \citet{aharonian2006} and \citet{takata2014}.  
Also the previous \fermi analyses cover at most 4 years of data in \citet{takata2014}. 
Here by taking all available data that cover more than 10 years of the \fermi observations, 
we have chance to look into the $\gamma$-ray spectrum in fine orbital phases. 
The strategy of our choosing the orbital phase is that, 
the entire phase is equally divided into four parts, 
where phases P and A cover symmetrically the orbital phases with respect to the periastron and apastron passages, 
while the the phases U and L are two orbital regions that are expected to have the same local emission environment.  
The advantages of having such subdivision of orbital phases is that 
we could disentangle the different contributions to the $\gamma$-ray emission 
via comparing with the properties  measured in the two identical regions U and L.
We derived the spectrum of these four parts respectively, as shown in Figure \ref{fig:sed_4bin},
while the fitting parameters with the likelihood analysis are listed in Table \ref{tab:sed_pars}.

Figure \ref{fig:sed_4bin} shows clearly an overall  evolution trend of the spectrum 
against the different orbital phases in an anti-clocked wise from P to L. 
The integral flux goes down from phase P,
accompanied with a harder spectral shape and larger cutoff energies.  
All the four spectra turn over at energies around 3--20\,GeV, 
which is consistent with the previous results in \citet{chang2016}.  
More details of the spectral fitting results are shown in Table \ref{tab:sed_pars}, 
where we see that for the two identical phase regions, 
the integral flux in phase U is larger than that in L, 
but the former has a relatively harder spectral shape and smaller cutoff energy,
shown in the right inset panel of Figure \ref{fig:sed_4bin}. 
A similar trend is also visible for the regions of P and A,
except that the phase P has softer spectral shape than phase A.

\begin{table}[!h]
\centering
\begin{tabular}{c|cccc}
\hline
Orbital Phase     & Energy Flux        &  Index           & Cut-off Energy   & TS  \\
        Range     &(10$^{-10}$\ergcms) &                  & (GeV)            &      \\
\hline
P: 0.00$\pm$0.125 &  4.138$\pm$0.003   & 2.435$\pm$0.001  & 3.682$\pm$0.008  & 9396  \\
U: 0.25$\pm$0.125 &  2.608$\pm$0.076   & 2.264$\pm$0.041  & 5.438$\pm$0.901  & 4187 \\
A: 0.50$\pm$0.125 &  1.902$\pm$0.072   & 2.268$\pm$0.057  & 7.168$\pm$1.759  & 2305 \\
L: 0.75$\pm$0.125 &  1.960$\pm$0.073   & 2.371$\pm$0.030  &12.372$\pm$2.937  & 2277 \\
\hline
\end{tabular}
\caption{Spectral parameters derived in the likelihood analysis.}
\label{tab:sed_pars}
\end{table}

\begin{figure}
\centering
\includegraphics[width=0.75\textwidth]{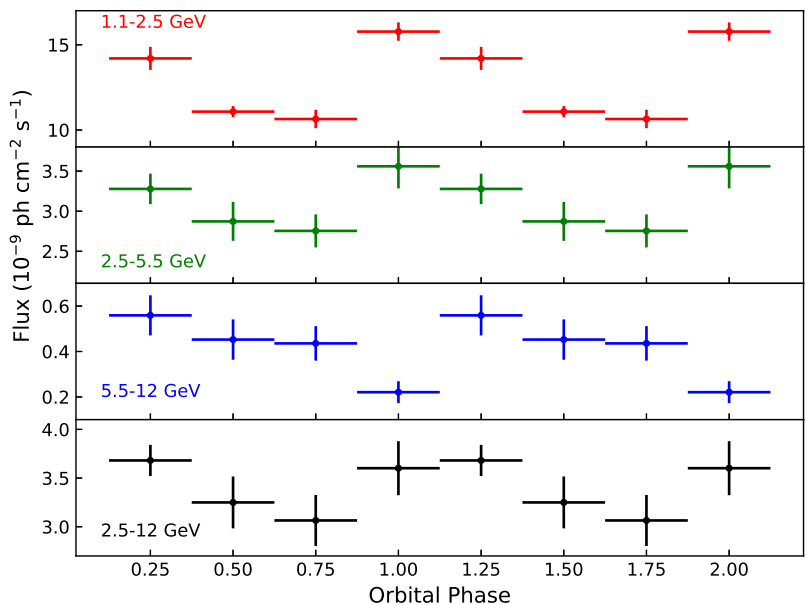}
\caption{Orbital light curves at energy band near the cut-off energy in our analysis.
A combination of the two light curves in 2.5--12\,GeV is shown in the $bottom$ panel.}
\label{fig:lc_3e}
\end{figure}

Orbital light curves at three different energy bands
(1.1--2.5\,GeV, 2.5--5.5\,GeV, 5.5-12\,GeV) around the cut-off energies 
listed in Table \ref{tab:sed_pars} are also shown in Figure \ref{fig:lc_3e}.
The modulation factors (($\rm Flux_{max}-Flux_{min})/(Flux_{max}+Flux_{min}$))
of the three light curves are 0.194$\pm$0.025, 0.128$\pm$0.053 and 0.434$\pm$0.109, respectively.
We can see that the orbital modulation is significantly lower at energies around the cut-off 
than at other energy bands, see Figure 5 in \citet{chang2016}.
A combination of the two light curves in 2.5--12\,GeV clearly shows that the orbital modulation is smeared out
(bottom panel of Figure \ref{fig:lc_3e}), with a modulation factor reduced to about 0.091$\pm$0.048.

\section{Discussion}
\label{sect:discussion}

We diagnose in detail the orbital phase dependence of  spectrum 
by adopting an optimized phase-interval definition and the most updated \fermi data that 
cover the last 10 years, which is 2.5 times more than that were taken in the previous reports \citep{takata2014}.  
Our results show obvious orbital phase evolution of the energy spectrum, 
and by taking the optimized phase-interval definition, 
these results may provide more clues to our better understanding of the emission mechanism in 
$\gamma$-ray binary like LS 5039. 

The energy spectrum derived for the four orbital phases can be described as a cut-off power law shape, 
and the trend that the spectrum at phase P is in general softer and the integral flux is higher than that in phase A 
is consistent with those reported before in SUPC and INFC \citep{collmar2014, chang2016}. 
Here, by taking more than 10 years \fermi data and an optimized phase-interval definition, 
the phases A and P spectra can be investigated by 
putting into a context of the spectral properties of other two identical phase regions U and L. 
Consequently, the impacts upon the energy spectrum may be disentangled: 
the phases U and L have the same local acceleration environment 
and supposed to have the same energy distribution for the relativistic plasma, 
hence the influences upon the energy spectrum are only about the stellar environment
once the relativistic particle moves out of the acceleration region; 
while in phases A and P an additional difference in the acceleration region 
due to the difference in distance to the companion has to be accounted.

Our previous results based on the timing analysis of LS 5039 revealed that 
at 3--20\,GeV the orbital modulation disappears \citep{chang2016}. 
Here our phase-resolved orbital  spectral analysis indicates that
all of the four individual phases have the similar spectral cut-off energy concentrating around 3--20\,GeV band, where the orbital modulation is largely smeared out. 

Supposing that the compact object of LS 5039 system is a neutron star, 
it is generally believed that  a shock can be formed via the collision 
between the stellar wind and the pulsar wind \citep{maraschi1981}. 
The shock will show up with a bow shape, 
in which the plasma can be  efficiently accelerated in the head of the shock 
and once the cooling plasma moves outward confined in the shock tails, 
the Doppler effect may boost the emission.  
The latter was once used by \citet{kong2012} to account for the \fermi flares of PSR B1259-63 
which occurred  after its periastron passage \citep{chang2018}.
Although the properties of LS 5039 is so far largely unknown, 
if we put LS 5039 in a context similar to PSR B1259-63, 
we may have the shock tails in phase U  with orientations almost opposite to the observer, 
while the shock tails in phase L region pointing toward  observer.
If the \fermi emissions are dominated by Doppler effect of the shock tails, 
it will then expect that the integral flux of phase U should be smaller than that in phase L. 
However, as shown in Table \ref{tab:sed_pars}, 
this is opposite to what we obtained from \fermi observations. 
Hence the \fermi emission should be dominated by inverse Compton scatterings of 
the particles accelerated in the shock head 
unless the shock configuration of LS 5039 is highly different from that of PSR B1259-63.
For the symmetric regions U and L, 
the  relativistic plasma released from the region U will go through more radiation field of the companion
and the collisions between the relativistic particles with the radiation field of the companion 
have larger incident angles in U than in L.
The inverse Compton effect will enhance the overall flux with respect to that in phase L 
resulting in lower cutoff energies in U.
This is consistent with what we measured for the spectral properties in phase U and phase L: 
the integral flux of phase U is higher while its cutoff energies are lower than those in phase L. 

As for the phases A and P, 
apart from the differences in stellar radiation environment that the accelerated particles have to experience, 
an additional difference in acceleration regions has to be considered. 
The stellar wind in the periastron region is denser than that in apastron. 
Therefore, the adiabatic cooling effect is stronger in periastron (Figure 6 in \citet{takahashi2009}) and 
prevent the particles from being accelerated to higher energies. 
As a result, the $\gamma$-ray spectrum in periastron should have a cutoff energy smaller than in apastron. 
This is in general consistent with the trend of what we measured. 
However, as shown in Table \ref{tab:sed_pars}, the cutoff energy ratio between phases A and P is about 1.95,
which is comparable with that 2.27 derived from the ratio between the phases L and U. 
The integral flux ratio of the former is about 2.18, significantly larger than that 1.33 of the latter. 
For the spectral shape, 
instead of having a spectral hardening with index decreasing from 2.37 in phase L to 2.26 in phase U, 
the spectral index is 2.27 in phase A but increase to 2.44 in phase P. 
These differences may be highly correlated to the change in properties of the local acceleration site: 
a denser stellar wind region like in phase P tend to produce via shocking the relativistic particles 
with energy spectrum characterized with softer spectrum shape and higher flux level. 
However, we notice that, the ratio of cutoff energy between U and L is comparable to that between P and A. 
This may indicate that the difference in particle energy spectrum produced via shock locally 
may not dominate the difference showing up in the high energy cutoff of the $\gamma$-ray spectrum, 
since the local particle spectra accelerated in regions U and L are supposed to be same. 
If this is true, the huge difference in TeV emission between periastron and apastron 
may not be due to the adiabatic cooling effect which results in a largely different spectrum of the accelerated particles,  
and hence the orbital modulation of TeV emission  should be from photon–photon absorption \citep{bottcher2007}
and  anisotropic IC scattering \citep{khangulyan2005}.

In summary, 
a joint diagnostic of the spectrum from symmetric phase of the orbit can provide more information
which turns out to be helpful for disentangling the impacts upon the $\gamma$-ray spectrum from different origins. 
As shown in our analysis of LS 5039 with more than 10 years of \fermi observations, 
the influences upon the $\gamma$-ray spectrum from the stellar radiation field 
can be clearly seen from two symmetric phases U and L, 
which have identical local environment for accelerating the particle via shock 
and hence in the two region the relativistic particles should have the same local spectral properties. 
In comparison with phases U and L, 
the change in local stellar wind density may result in different plasma spectrum that is intrinsic to the shock regions, 
which in turn is responsible to most differences
showing up in the emerged $\gamma$-ray spectrum in the phases A and P. 

\begin{acknowledgements}
The authors thank support from the National Key R\&D Program of China (Grant No. 2016YFA0400800), 
the NSFC (Nos. U1838201, U1838202 and 11733009), XTP project (XDA 04060604)
and the Strategic Priority Research Programme `The Emergence of Cosmological Structures'
of the Chinese Academy of Sciences (Grant No.XDB09000000).
\end{acknowledgements}

\label{lastpage}


\begin{thebibliography}{99}
\bibliographystyle{plain}
\bibitem[Abdo et al.(2009)]{abdo2009}
	Abdo A. A., Ackermann M., Ajello M., et al., 2009, \apj, 706, L56

\bibitem[Aharonian et al.(2006)]{aharonian2006}
	Aharonian F., Akhperjanian A. G., Bazer-Bachi A. R., et al., 2006, \aap, 460, 743 

\bibitem[Atwood et al.(2009)]{atwood2009} 
	Atwood W. B., Abdo A. A., Ackermann M., et al., 2009, \apj, 697, 1071 

\bibitem[B\"ottcher et al.(2007)]{bottcher2007}
	B\"ottcher, M., \& Dermer, C. D. 2007, Astrophys Space Sci, 307, 233

\bibitem[Casares et al.(2005)]{casares2005}
	Casares J., Ribo M., Ribas I., et al., 2005, \mnras, 364, 899 

\bibitem[Chang et al.(2016)]{chang2016}
	Chang Z., Zhang S., Ji L., et al.,  2016, \mnras, 463, 495 

\bibitem[Chang et al.(2018)]{chang2018}
	Chang Z., Zhang S., Chen Y. P., et al., 2018, \raa, 18, 152

\bibitem[Collmar et al.(2014)]{collmar2014}
	Collmar W., \& Zhang S., 2014, \aap, 565, A38

\bibitem[Dubus(2013)]{dubus2013} 
	Dubus G., 2013, \aapr, 21, 64

\bibitem[Dubus et al.(2015)]{dubus2015} 
	Dubus G., Lamberts A., Fromang S., 2015, \aap, 581, A27

\bibitem[Hadasch(2012)]{hadasch2012}
	Hadasch D., Torres D. F., Tanaka T., et al., 2012, \apj, 749, 54

\bibitem[Johnston et al.(1992)]{johnston1992}
	Johnston S., Wood K. S., Ray P. S., et al.,1992, \apj, 387, L37

\bibitem[Khangulyan et al.(2005)]{khangulyan2005}
	Khangulyan, D., \& Aharonian, F. 2005, AIP Conference Proceedings, Vol. 745, 359

 \bibitem[Kong et al.(2012)]{kong2012} 
 	Kong S. W., Cheng K. S., \& Huang Y. F. 2012, \apj, 753, 127

\bibitem[Maraschi et al.(1981)]{maraschi1981} 
	Maraschi L., \& Treves A., 1981, \mnras, 194, 1P

\bibitem[Mattox et al.(1996)]{mattox1996} 
	Mattox J. R., Bertsch D. L., Chiang J., et al., 1996, \apj, 461, 396

\bibitem[Takahashi et al.(2009)]{takahashi2009}
	Takahashi T., Kishishita T., Uchiyama Y., et al., 2009, \apj, 697, 592

\bibitem[Takata et al.(2014)]{takata2014}
	Takata J., Leung G. C. K., Tam P. H. T., et al., 2014, \apj, 790, 18

\end{thebibliography}
\end{document}